\documentclass[12pt]{article}
\usepackage{amsmath,amsfonts,amssymb,geometry}
\usepackage{graphicx,indentfirst,verbatim,epsfig,afterpage,caption}
\usepackage[onehalfspacing]{setspace}
\usepackage{graphicx}
\usepackage{accents}
\usepackage{booktabs}
\usepackage{url}
\usepackage{amsmath}
\usepackage{graphicx}
\usepackage{hyperref}
\usepackage{caption}
\usepackage{booktabs}
\usepackage{geometry}
\usepackage{setspace}
\usepackage{float}  % Add this line in the preamble

\usepackage{adjustbox} % Added for adjusting table size

\begin{document}
\renewcommand{\thetable}{\Roman{table}}

\title{Behavioral Expectations in New Keynesian DSGE Models: Evidence from India's COVID-19 Recovery and Vaccination Program}
\author{Arpan Chakraborty\thanks{Corresponding author. PhD Scholar, Department of Humanities and Social Sciences, Indian Institute of Technology Kharagpur, Kharagpur, West Bengal, India. arpan.ms97@kgpian.iitkgp.ac.in, ORCID - 0000-0002-7777-5643} 
\and Siddhartha
Chattopadhyay\thanks{Associate Professor, Department of Humanities and
Social Sciences, Indian Institute of Technology Kharagpur, Kharagpur, West Bengal, India.
siddhartha@hss.iitkgp.ac.in, ORCID - 0000-0001-8663-0246} }
% \and Sohini Sahu%
% \thanks{%
% Indian Institute of Technology Kanpur, Department of Economic Sciences,
% Kanpur, Uttar Pradesh, India. ssahu@iitk.ac.in, ORCID - 0000-0001-7293-5671}}
\date{ \today }
\maketitle

\maketitle

%Suppress page numbering on the title page
\thispagestyle{empty}

%Start a new page for the abstract, keywords, and JEL codes
\newpage

%Suppress page numbering for this page as well
\thispagestyle{empty}

\begin{center}
\textbf{Abstract:}
\end{center}

\textbf{Purpose --}  
This paper extends the New Keynesian Dynamic Stochastic General Equilibrium (DSGE) framework by incorporating behavioral expectations to analyze India's post-pandemic output gap and inflation rate moments, focusing on the effects of COVID-19 and vaccination. By departing from rational expectations models, we show that behavioral expectations better capture the distributional characteristics of India's output gap and inflation rate.

\textbf{Design/methodology/approach --}  
Following Dasgupta and Rajeev (2023), we model the pandemic as a prolonged negative demand shock and vaccinations as a positive supply shock, with the initial negative demand shock derived from observed declines in the output gap. We then identify the optimal positive supply shock and persistence parameters of both shocks by minimizing the Mahalanobis distance, aligning with the averages of the filtered output gap data and the inflation rate data. India's output gap is estimated using both HP and Kalman filter techniques.

\textbf{Findings --}  
Using Mahalanobis distance minimization, we obtain the optimal vector comprising the initial positive supply shock and the persistence parameters of the negative demand and positive supply shocks. The calibrated values for the models, corresponding to the actual output gap contractions derived from the Hodrick-Prescott and Kalman filters, are (0.64, 0.8, 0.9) and (0.57, 0.8, 0.95), respectively. Additionally, we extend the work of Dasgupta and Rajeev (2023), who had presented only a static simple Keynesian model, by incorporating a DSGE framework.

\textbf{Originality/value --}  
This study makes a significant contribution to the literature in the following ways:  

\begin{enumerate}
    \item The first paper in the literature to introduce behavioral expectations within a DSGE framework to analyze India's output gap and inflation rate during COVID-19.
    \item It demonstrates that behavioral expectations models are more effective than rational expectations in capturing the distributional characteristics of India's output gap and inflation rate. Also, we quantitatively extend previous work by Dasgupta and Rajeev (2023) by providing a New Keynesian DSGE model.
\end{enumerate}

\textbf{Paper type:} Research paper



\pagebreak

% \thispagestyle{empty}%

% \end{abstract}

\pagebreak
\setcounter{page}{1}

\section{Introduction}
The Dynamic Stochastic General Equilibrium (DSGE) model has become a cornerstone of modern macroeconomic analysis, particularly in understanding business cycles in developed economies. Despite its widespread use, its application to emerging economies like India remains limited\footnote{Peiris, Saxegaard, and Anand (2010) developed a DSGE model for India with macro-finance linkages, using Bayesian estimation on 1996–2008 data. Subsequent studies explored specific objectives: Banerjee, Basu, and Ghate (2020) analyzed weak monetary transmission; Sarkar (2020, 2022) examined stock markets, COVID-19 impacts, and monetary transfers; Arora et al. (2024) examined the sectoral impact of COVID-19 on the Indian economy using a GTAP model; Shah and Garg (2023) assessed post-pandemic policy responses; Sharma and Behera (2022) highlighted DSGE models' superiority in output gap analysis; and Kumar (2023) studied productivity and monetary policy shocks, showing their distinct effects on growth and output gaps. In the empirical field, Ali and Khan (2024) provided empirical evidence from rural India, showing that sociodemographic and institutional factors significantly influenced access to food during the COVID-19 lockdown.}. Dasgupta and Rajeev (2020, 2023) have emphasized the lack of macro-theoretical research focused on India, especially in the context of the COVID-19 pandemic. Their work utilized a simple Keynesian framework to analyze the pandemic's effects, incorporating negative demand shocks induced by COVID-19 and vaccination-driven positive supply shocks simultaneously. However, their analysis was primarily based on a static model, and they did not employ New Keynesian DSGE modeling.

To address this research gap, our study introduces a behavioral New Keynesian DSGE model tailored to India's economic dynamics during the pandemic. We model this using behavioral expectation formation as it better matches the moments of the output gap data compared to rational expectation formation for the US economy (See De Grauwe, 2012; De Grauwe and Ji, 2019, 2020; Chakraborty et al., 2024).\footnote{Chakraborty et al. (2024) further demonstrate that behavioral expectation formation also aligns well with the duration of long-term bonds for the US.}

Our study makes the following distinct contributions to the existing literature. First, we show that the simple behavioral New Keynesian DSGE model matches the output gap and inflation rate moments better than the rational expectation DSGE model in the post-pandemic era.\footnote{Note that, for simplicity, we take parameters from the existing literature.}

Second, we extend this framework to analyze the economic impact of India's COVID-19 vaccination program. We offer a more rigorous quantitative assessment based on Dasgupta and Rajeev's (2023) work, which provided a graphical analysis using a simple Keynesian model. We modeled the pandemic as a combination of deterministic autoregressive negative demand shock and positive supply shock, setting the magnitude of the initial shock equal to the observed output gap dip from the HP and Kalman filters. We then calibrate the initial vaccination-induced value of the positive supply shock and the persistence parameters of the negative demand and positive supply shocks. Our simulations also closely match the actual moments of the output gap and the inflation rate.

Our analyses reveal that the optimal vector comprising the initial positive supply shock and the persistence parameters of the negative demand and positive supply shocks takes calibrated values of (0.64, 0.8, 0.9) and (0.57, 0.8, 0.95) for models employing initial negative demand shock corresponding to the actual output gap contractions derived from the Hodrick-Prescott and Kalman filters, respectively. We utilize minimum Mahalanobis distance as the criterion to determine this optimal vector of values. Unlike other distance-minimizing approaches, we employ Mahalanobis distance, as it is scale-invariant and accounts for the covariance structure between the actual and simulated data moments.

This research is particularly timely, given ongoing debates about policy responses to economic crises after the global pandemic. By demonstrating the superior fit of behavioral expectations in capturing Indian output gap dynamics, we provide policymakers with a more accurate framework for decision-making. Furthermore, our quantification of vaccination impacts offers valuable information for future integration of public health and economic policy.

The rest of the paper is organized as follows: Section 2 describes the data, Section 3 introduces the model, Section 4 depicts the results, and finally, Section 5 concludes.

\section{Data}
The primary challenge faced by this paper lies in data collection, particularly for the output gap of the Indian economy. Unlike developed economies, real output gap data for India is not directly available and is typically estimated using various econometric techniques. This paper adopts a similar approach\footnote{Salunkhe and Patnaik (2019) adopted a similar approach to calculate the output gap of the Indian economy using the Hodrick Prescott filter.}. We obtain the Quarterly real GDP data for the Indian economy from FRED\footnote{https://fred.stlouisfed.org/series/NGDPRNSAXDCINQ}. Using this dataset\footnote{The output gap data ranges from 1st quarter 2004 to the 1st quarter 2024. For our COVID related analysis, we use data only after Q1 2020.}, we calculate the output gap using HP and Kalman filter. Similarly, we sourced the quarterly CPI data from FRED and calculated the quarter-on-quarter inflation rate\footnote{We adjust the CPI data to ensure that both datasets share the same base year, with Q4 2011 serving as the base year for both.}.

\section{The Model}
As described in the introduction, this paper uses the three equation behavioral New Keynesian (NK) macroeconomic framework developed by De Grauwe (2012), De Grauwe and Ji (2019) in the context of the Indian economy. The Aggregate Demand (AD) equation is as follows:

\begin{equation}
y_{t}=\widetilde{E}_{t}(y_{t+1})-\frac{1}{\sigma}(i_{t}-\widetilde{E}_{t}(\pi _{t+1}))+\epsilon _{t};\quad t=1,2,3,... 
\label{1}
\end{equation}
where, $y_{t}$ denotes the output gap, $i_{t}$ is the short-term nominal interest rate. The expectations in this model are non-rational.

The Phillips curve/AS equation, derived under monopolistic competition and Calvo pricing (Calvo, 1983), is as follows:

\begin{equation}
\pi _{t}=\beta \widetilde{E}_{t}(\pi _{t+1})+\kappa y_{t}+\eta_{t}
\label{2}
\end{equation}
where, $0\leq \kappa\leq 1$ measures the sensitivity of inflation to the output gap, $\beta$ measures the discount rate, and $\pi _{t}$ is the inflation rate. Also, similar to the AD shock, we introduce the AS shock $\eta _{t}$. Following De Grauwe and Ji (2020), the parameter $\kappa$ can be expressed as:

\begin{equation*}
\kappa = \frac{(1-\theta)(1-\beta \theta)}{\theta} \frac{\sigma (1-\varsigma) + \chi + \varsigma}{1-\varsigma + \varsigma \acute{e}}
\end{equation*}
where $1-\theta$ reflects the expected price of the Calvo lottery ticket, $\chi$ denotes the inverse Frisch elasticity of labor supply in the household utility function, $\varsigma$ represents the labor elasticity in the monopolistically competitive labor-augmented production function, $[Y_{t}^{i} = A_{t}L_{t}^{1-\varsigma ,i}]$, and $\acute{e}$ is the price elasticity of demand, which determines the markup price, $M$, for monopolistically competitive firms, $[M = \frac{\acute{e}}{\acute{e}-1}]$ (see Gali, 2008). 

Note a $\theta$ fraction of firms cannot adjust prices in each period. For $\theta = 1$, $b_{2}$ equals zero, indicating highly rigid prices, while for $\theta = 0$, $b_{2}$ approaches infinity.

The short-term nominal interest rate ($i_{t}$) is set according to the Taylor rule\footnote{While studies like Chattopadhyay (2013) incorporate a zero lower bound constraint in his monetary policy rule specifications, we abstain from this restriction for Indian economy.} (Taylor, 1993; Blattner and Margaritov, 2010):

\begin{equation}
i_{t} = (1-c_{3})(c_{1}\pi _{t} + c_{2}y_{t}) + c_{3}i_{t-1}
\label{3}
\end{equation}
where $c_{1}>1$, $0<c_{2}<1$ are the coefficients governing inflation and output gap responses, respectively. Note that the term $c_{3}$ captures interest rate smoothing.

\subsection{Simulation Parameters}
Table \ref{tab:simulation_parameters} lists the simulation parameters, aligned with the papers described below:
\begin{table}[h!]
    \centering
    \caption{Simulation Parameters for the Indian Economy}
    \small % Reduce font size for table content
    \renewcommand{\arraystretch}{1.1} % Adjust row height
    \begin{tabular}{|l|p{4.6in}|} % Define column width to fit page
        \hline
        \textbf{Parameter} & \textbf{Description and Source} \\
        \hline
        \( \kappa = 0.065 \) & Output gap coefficient in AS (Calculated using the formula) \\
        % \hline
        \( \beta = 0.98 \) & Discount factor (Gabriel et al., 2012) \\
        % \hline
        \( \acute{e} = 7.01 \) & Price elasticity of demand (Gabriel et al., 2012) \\
        % \hline
        \( \sigma = 1.5 \) & CRRA of household consumption (Das and Nath, 2019; Gabriel et al., 2012) \\
        % \hline
        \( \varsigma = 0.7 \) & Share of labor in production function (Banerjee, Basu, and Ghate, 2020) \\
        % \hline
        \( \chi = 2.7 \) & Inverse of Frisch elasticity of labor supply (Anand and Prasad, 2010; Sharma and Behera, 2022) \\
        % \hline
        \( \theta = 0.75 \) & Calvo price rigidity (Kumar, 2023) \\
        % \hline
        \( c_{1} = 1.2 \) & Interest rate sensitivity of inflation (Gabriel et al., 2012) \\
        % \hline
        \( c_2 = 0.5 \) & Interest rate sensitivity of output (Banerjee, Basu, and Ghate, 2020) \\
        % \hline
        \( c_3 = 0.8 \) & Interest rate smoothing parameter (Banerjee, Basu, and Ghate, 2020) \\
        \( \gamma = 2 \) & Learning intensity (De Grauwe and Ji, 2019) \\
        \( \rho = 0.5 \) & Memory parameter (De Grauwe and Ji, 2019) \\
        \( T = 2000 \) & Simulation length \\
        \hline
    \end{tabular}
    \label{tab:simulation_parameters}
\end{table}

It is important to note that this study does not have the calibrated values for the memory parameter and the willingness-to-learn parameter specific to the Indian economy. Therefore, for simplicity, this paper uses the parameters given by De Grauwe and Ji (2019). Furthermore, unlike Kumar (2023), this article incorporates aggregate supply (AS) shocks into the analysis to analyze the impact of vaccination. Please refer to the Appendix for more details on the HP and Kalman filter techniques to compute the output gap data.

\subsection{Behavioral Expectation Formation}
% Following De Grauwe and Ji (2019), the behavioral expectations model includes fundamentalist and extrapolator agents. The expectation of the output gap \((y_t\)) is as follows:

% \[
% \widetilde{E}_{t}(y_{t+1}) = \alpha_{f,t}^{y}\widetilde{E}_{t}^{f}(y_{t+1}) + \alpha_{e,t}^{y}\widetilde{E}_{t}^{e}(y_{t+1})
% \]
% where fundamentalists expect \(\widetilde{E}_{t}^{f}(y_{t+1}) = y_{ss}\), \footnote{which is normalized to zero.} and extrapolators use \(\widetilde{E}_{t}^{e}(y_{t+1}) = y_{t-1}\), and $\alpha_{f,t}^{y}$ and $\alpha_{e,t}^{y}$ are the proportions based on the relative forecast performance (mean square forecast errors), which is as follows:

% \[
% U_{f,t} = -\sum_{k=0}^{\infty} \varrho_k \left[y_{t-k-1} - \widetilde{E}_{f,t-k-2}(y_{t-k-1})\right]^2
% \]

% \[
% U_{e,t} = -\sum_{k=0}^{\infty} \varrho_k \left[y_{t-k-1} - \widetilde{E}_{e,t-k-2}(y_{t-k-1})\right]^2
% \]
% where $\varrho_k = (1-\rho)\rho^k$, with $0 < \rho < 1$ as the memory parameter. Hence, the proportion of fundamentalists becomes:

% \[
% \alpha_{f,t}^{y} = \frac{\exp(\gamma U_{f,t})}{\exp(\gamma U_{f,t}) + \exp(\gamma U_{e,t})}, \quad \alpha_{e,t}^{y} = 1 - \alpha_{f,t}^{y}
% \]
% where $\gamma$ is the willingness to learn/intensity of choice parameter;
% \subsection{Behavioral Expectations and Forecasting Rule Selection}

We use the behavioral expectation model of De Grauwe (2012) for our analysis. In this type of behavioral expectations model, agents use different strategies to form expectations about future output gaps \( y_t \). Here, we have two types of agents: fundamentalists and extrapolators. Fundamentalists expect the output gap to return to its steady state \( y_{ss} \), while extrapolators predict future values based on the past output gap. The model combines their expectations as follows:

\[
\widetilde{E}_t(y_{t+1}) = \alpha_{f,t}^y \widetilde{E}_t^f(y_{t+1}) + \alpha_{e,t}^y \widetilde{E}_t^e(y_{t+1}),
\]
where \( \alpha_{f,t}^y \) and \( \alpha_{e,t}^y \) are the proportions of fundamentalists and extrapolators. These proportions depend on the relative forecast performance of the two agents, calculated using their mean squared forecast errors (MSFEs).

\subsubsection{Forecast Performance and Utility Functions}

The deterministic utility functions \( \widehat{U}_{f,t} \) for fundamentalists and \( \widehat{U}_{e,t} \) for extrapolators are based on the mean squared forecast errors between their forecasts and actual outcomes over time. The utilities are:

\[
\widehat{U}_{f,t} = -\sum_{k=0}^{\infty} \varrho_k \left[ y_{t-k-1} - \widetilde{E}_{f,t-k-2}(y_{t-k-1}) \right]^2,
\]
\[
\widehat{U}_{e,t} = -\sum_{k=0}^{\infty} \varrho_k \left[ y_{t-k-1} - \widetilde{E}_{e,t-k-2}(y_{t-k-1}) \right]^2.
\]

Here, \( \varrho_k = (1-\rho)\rho^k \) assigns geometrically declining weights to past errors, with \( \rho \) as the memory parameter (\( 0 < \rho < 1 \)). The decreasing weights reflect the idea that agents pay more attention to recent errors than to past ones. Agents are assumed to forget past errors over time.

\subsubsection{Discrete Choice Theory and Rule Selection}

Agents need to decide which forecasting rule to use. We model this decision process using discrete choice theory, as described by Brock and Hommes (1997, 1998). Agents compare the forecast performance (utility) of the two rules, but they are also influenced by random factors like their mood or external conditions. The utilities of the rules are:

\[
U_{f,t} = \widehat{U}_{f,t} + e_{f,t}, \quad U_{e,t} = \widehat{U}_{e,t} + e_{e,t},
\]
where \( \widehat{U}_{f,t} \) and \( \widehat{U}_{e,t} \) are the deterministic utility components, and \( e_{f,t} \) and \( e_{e,t} \) are random components. These random components are assumed to follow a logistic distribution\footnote{$\gamma$ is the precision of the distribution.}. The probability that an agent chooses the fundamentalist rule is:

\[
P(\text{fundamentalist})=\alpha_{f,t}^{y} = \frac{\exp(\gamma U_{f,t})}{\exp(\gamma U_{f,t}) + \exp(\gamma U_{e,t})}, \quad \alpha_{e,t}^{y} = 1 - \alpha_{f,t}^{y}
\]
where $\alpha_{f,t}^{y}$ and $\alpha_{e,t}^{y}$ are the proportions based on the relative forecast performance (mean square forecast errors), and they represent the proportion/probabilities of fundamentalists and extrapolators,, respectively. This probability increases when the forecast performance of the fundamentalist rule is better than the extrapolator. The parameter \( \gamma \) determines how much agents learn from past mistakes in forecasting. Or in other words, it represents the willingness to learn/intensity of choice parameter.

Agents in the model continuously learn from their past mistakes and adapt their behavior. If agents see that their chosen rule is underperforming, they will switch to the other rule. This learning process helps agents avoid repeating errors. The model assumes that agents have bounded rationality, meaning they do not have perfect knowledge of the economy. Instead, they use simple and parsimonious forecasting rules and adjust them based on experience.

This type of model contrasts with mainstream models, which assume that agents can predict the future and make optimal decisions. This model provides a more realistic view of how expectations form and change over time by incorporating learning and adaptation.

We also determine the simulated inflation rate using independent and identical expectation formation. For programming, we use the Binder and Pesaran (2000) technique and the Matlab code of De Grauwe (2012), to simulate the rational and the behavioral expectation DSGE model.

\subsection{Structural Break in Output Gap}
To check for the structural break, we plot the series of output gap data below. Figure 1 reveals a visible structural break\footnote{Kadiyala and Ascioglu (2024) find that COVID-19 lockdowns caused a default surge to 95.29\% in April 2020. We find a similar decline in the output gap in Q1 2020.} in \textbf{Q1 2020}. 

\begin{figure}[H]
    \centering
    \includegraphics[width=\textwidth]{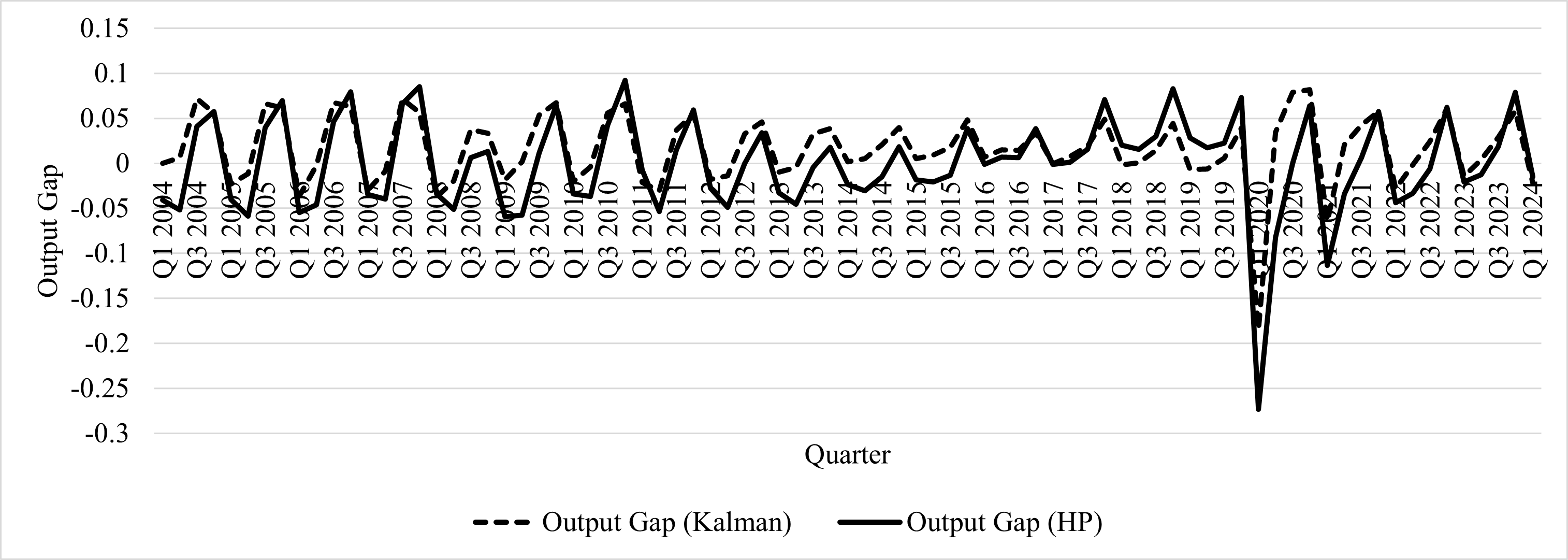} % Replace with your file name and adjust width as needed
    \caption{Quarterly Output Gap of India: Comparison of HP and Kalman Filter} % Replace with your title
    \label{fig:example_figure} % Optional: for referencing the figure
\end{figure}

To formally validate this observation, we employ a \textit{Likelihood Ratio (LR) test} to examine the presence of a structural break in the Output Gap series, estimated using both Kalman and HP filters. Please refer to the Appendix for more details. Similarly, we plot the quarterly inflation rate in Figure 2 below.

\begin{figure}[h!]
    \centering
    \includegraphics[width=\textwidth]{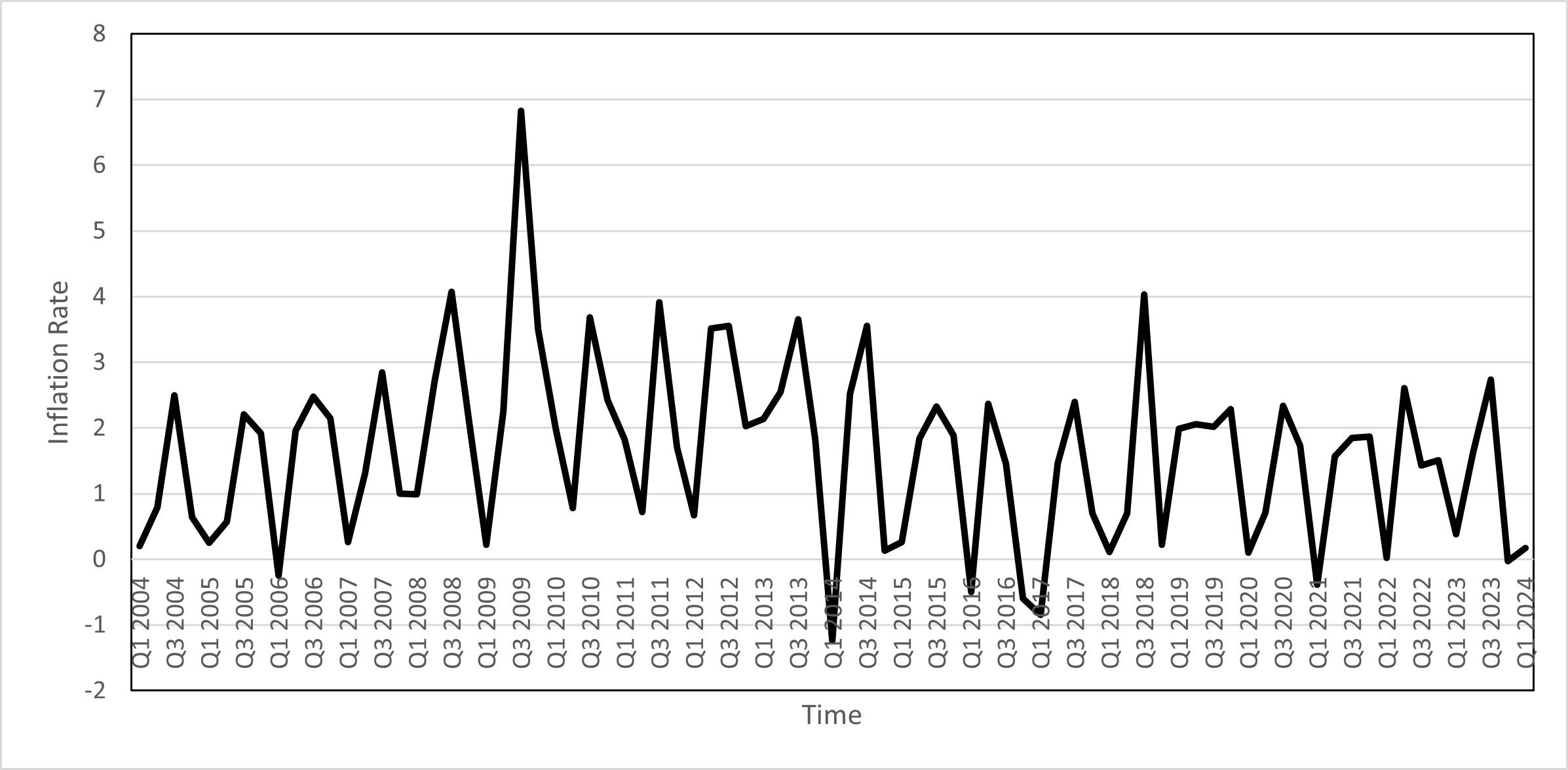} % Replace with your file name and adjust width as needed
    \caption{Quarterly Inflation rate of India using CPI data} % Replace with your title
    \label{fig:example_figure1} % Optional: for referencing the figure
\end{figure}

\section{Results}
Building upon the existing literature corroborating the superior performance of behavioral expectations and the verified Q1 2020 structural break in India's output gap, we analyze the COVID-19-related policy implications within our DSGE model.

Our primary objective is to determine the initial magnitude of the positive supply shock (\(\eta_1\)) precipitated by vaccination implementation, and the persistence parameters of negative aggregate demand (\(\rho_\epsilon\)) and positive aggregate supply (\(\rho_\eta\)) shocks to achieve correspondence with post-COVID actual average values of the output gap and inflation rate. To do this, we conduct a \textbf{grid search} over (\(\eta_1\)), (\(\rho_\epsilon\)), and (\(\rho_\eta\)) ranging from 1 to 0 in increments of 0.01, 0.05, and 0.05, respectively\footnote{Out of each run, we utilize only 16 quarters of simulations out of 2000 to match the moments of the available empirical data. In total, we ran 88.2 million simulations for each model.}.

We first introduce a \textbf{negative aggregate demand shock} (\(\epsilon_t\)), spanning ten quarters\footnote{As COVID-19 lasted for 10 quarters in India.}, with an initial magnitude of \(\epsilon_1 = -0.27\) (HP-filtered output gap for Q1 2020). This represents \(\epsilon_1\), the initial value of the COVID shock in our model. 

Moreover, we assume that the shock follows an autoregressive decay (\(\rho_\epsilon\)), modeled as:

\begin{equation}
\epsilon_t = \rho_\epsilon^{(t-1)} \cdot \epsilon_1
\end{equation}
where \(\rho_\epsilon\) captures the decay rate of COVID-19 pandemic. The initial shock values are given by \(\epsilon_1 = -0.27\) for the HP-filtered output gap and \(\epsilon_1 = -0.18\) for the Kalman-filtered output gap. Note that, we simulate two different models.

Similarly, to model the impact of vaccinations, we introduce a \textbf{positive aggregate supply shock} (\(\eta_t\)) spanning six quarters\footnote{Vaccines were first made available in India in Q1 2021. We assume that vaccination efforts contributed to the economic recovery, effectively mitigating the crisis from Q1 2021 to Q2 2022, a period of six quarters. Consequently, the positive supply shock is applied only during this timeframe.},

\begin{equation}
    \eta_t = \rho_\eta^{(t-1)} \cdot \eta_1
\end{equation}
where the positive supply shock follows the autoregressive decay process as the aggregate demand shock, and \(\rho_\eta\) represents the persistence of the shock. The distinct values of \(\eta_1\) are calibrated separately for the HP-filter-based and Kalman-filter-based models.

The initial magnitude of the shock, \(\eta_1\), is unknown and is calibrated by matching the empirical post-pandemic output gap and the inflation rate moments (average) while minimizing the Mahalanobis distance (M.D.). The Mahalanobis distance \(d\) is given by:

\[
d = \sqrt{(\mathbf{s}_{\text{sim}} - \mathbf{s}_{\text{data}}) \cdot \Sigma^{-1} \cdot (\mathbf{s}_{\text{sim}} - \mathbf{s}_{\text{data}})^\top}
\]
where our methodology involves defining the simulated and actual vectors. The simulated vector, denoted by \(\mathbf{s}_{\text{sim}}\), is represented as a 2-dimensional vector containing the mean values of the simulated output gap and inflation rate:

\[
\mathbf{s}_{\text{sim}} = \begin{bmatrix} \text{mean}_1 & \text{mean}_2 \end{bmatrix}
\]
and the actual data vector, denoted by \(\mathbf{s}_{\text{data}}\), is defined as:

\[
\mathbf{s}_{\text{data}} = \begin{bmatrix} \text{mean}_{\text{Output gap}} & \text{mean}_{\text{inf}} \end{bmatrix}
\]
given this, we compute the covariance matrix \(\Sigma\) between the simulated and actual vectors. This matrix captures the relationship between the simulated and actual data, and is as follows:
\[
\Sigma = \text{cov}(\mathbf{s}_{\text{data}}, \mathbf{s}_{\text{sim}})
\]
Finally, by minimizing the Mahalanobis distance\footnote{Note that, Mahalanobis distance is unit-free. Also, Mahalanobis distance adjusts for dynamic correlations.}, and by calibrating the grid of persistence parameters \(\rho_{\epsilon}\), \(\rho_{\eta}\), and the initial positive supply shock \(\eta_1\), we obtain our results for both rational and behavioral expectations\footnote{We apply same shock to both models.}.

Table \ref{tab:hp_results} presents the results of the analysis based on the HP filter, comparing the models of rational and behavioral expectations. In this analysis, we focus on three key parameters as previously written. The results show that for the behavioral model, the persistence values of \(\rho_{\epsilon}\) and \(\rho_{\eta}\) are 0.8 and 0.9, respectively, with the initial supply shock value set to 0.64. In contrast, the rational expectation model assumes no persistence for the COVID shock (\(\rho_{\epsilon} = 0.0\)) and a persistence of 1.0 for the vaccination shock, with an initial supply shock value of 1.0.

The mean output gap and inflation rate for the behavioral model are -0.0035 and 1.2518, respectively, while for the rational expectation model, these values are significantly different at -0.20305 and 0.3618. This discrepancy is reflected in the Mahalanobis distance, which is smaller for the behavioral model (1.3794) compared to the rational model (1.4142), suggesting that the behavioral model better matches the empirical data. The actual mean values for the output gap and the inflation rate are -0.0046 and 1.2580, respectively, closely aligning with the results of the behavioral model.

\begin{table}[h]
\centering
\caption{HP-Filter Based Moments Comparison}
\small
\renewcommand{\arraystretch}{1.1}
\begin{tabular}{|l|c|c|}
\hline
\textbf{Parameter / Statistic} & \textbf{Behavioral} & \textbf{Rational} \\
\hline
\( \rho_{\epsilon} \) (Covid Shock Persistence) & 0.8 & 0.0 \\
\( \rho_{\eta} \) (Vaccination Persistence) & 0.9 & 1.0 \\
Initial Supply Shock & 0.64 & 1.0 \\
\hline
Mean (Output Gap) & -0.0035 & -0.20305 \\
Mean (Inflation Rate) & 1.2518 & 0.3618 \\
Mahalanobis Distance & 1.3794 & 1.4142 \\
\hline
Actual Mean (HP Output Gap) & -0.0046 & - \\
Actual Mean (Inflation Rate) & 1.2580 & - \\
\hline
\end{tabular}
\label{tab:hp_results}
\end{table}

Table \ref{tab:kalman_results} presents the results of the Kalman-filter based analysis, which follows a similar structure to the HP-filter model. For the behavioral model, the persistence values of \(\rho_{\epsilon}\) and \(\rho_{\eta}\) are 0.8 and 0.95, respectively, with the initial supply shock set to 0.57. The rational expectation model assumes the same persistence for the vaccination shock as in the HP-filter model (\(\rho_{\eta} = 1.0\)), with an initial supply shock value of 1.0.

The results show that the behavioral model for the Kalman filter yields a mean output gap of 0.027 and a mean inflation rate of 1.2557, while the rational expectation model again produces values significantly differing from the actual data: a mean output gap of -0.20421 and a mean inflation rate of 0.3617. The Mahalanobis distance for the behavioral model is 1.3628, closer to the empirical data than the rational model’s distance of 1.4142. The actual mean output gap and inflation rate for the Kalman filter are 0.0227 and 1.2580, respectively, confirming that the behavioral model provides a better fit.

\begin{table}[h]
\centering
\caption{Kalman-Filter Based Moments Comparison}
\small
\renewcommand{\arraystretch}{1.1}
\begin{tabular}{|l|c|c|}
\hline
\textbf{Parameter / Statistic} & \textbf{Behavioral} & \textbf{Rational} \\
\hline
\( \rho_{\epsilon} \) (Covid Shock Persistence) & 0.8 & 0.0 \\
\( \rho_{\eta} \) (Vaccination Persistence) & 0.95 & 1.0 \\
Initial Supply Shock & 0.57 & 1.0 \\
\hline
Mean (Output Gap) & 0.027 & -0.20421 \\
Mean (Inflation Rate) & 1.2557 & 0.3617 \\
Mahalanobis Distance & 1.3628 & 1.4142 \\
\hline
Actual Mean (Kalman Output Gap) & 0.023 & - \\
Actual Mean (Inflation Rate) & 1.2580 & - \\
\hline
\end{tabular}
\label{tab:kalman_results}
\end{table}

\subsection{Higher order Moments comparison}
In this analysis, we simulate the model using the optimal values of \(\rho_{\epsilon}\), \(\rho_{\eta}\), and \(\eta_1\) obtained from the behavioral expectations model for both the HP and Kalman-filter-based analysis.

In Table \ref{tab:hp_properties}, the simulated output gap has a mean of -0.0035, variance of 0.0609, skewness of 0.39, and kurtosis of 1.81. The simulated inflation rate shows a mean of 1.2518, variance of 0.5621, skewness of -0.81, and kurtosis of 2.10. The Jarque-Bera p-value for both the simulated output gap and the inflation rate is 0.22 and 0.09, respectively, indicating that the simulated data reasonably approximate the empirical distribution and they follow a normal distribution.

Similarly, in Table \ref{tab:kalman_properties} , the simulated output gap has a mean of 0.0273, variance of 0.0500, skewness of 0.60, and kurtosis of 2.01. The simulated inflation rate has a mean of 1.2557, variance of 0.5879, skewness of -0.72, and kurtosis of 1.95. The Jarque-Bera p-value for the simulated output gap and the inflation rate is 0.15 and 0.10, respectively, suggesting that both the simulated and the actual data for the Kalman filter also follow a normal distribution, which further supports the robustness of our analysis.

\begin{table}[h]
\centering
\caption{Statistical Properties of HP-Based Output Gap}
\small
\renewcommand{\arraystretch}{1.5}
\begin{tabular}{|l|c|c|c|c|c|}
\hline
\textbf{Data} & \textbf{Mean} & \textbf{Variance} & \textbf{Skewness} & \textbf{Kurtosis} & \textbf{JB-P Value} \\
\hline
HP (Actual) & -0.0046 & 0.0028 & -0.19 & 2.57 & 0.50 \\
Inflation (Actual) & 1.2580 & 0.9721 & -0.22 & 1.83 & 0.35 \\
Simul Inflation & 1.2518 & 0.5621 & -0.81 & 2.10 & 0.09 \\
Simul Output & -0.0035 & 0.0609 & 0.39 & 1.81 & 0.22 \\
\hline
\end{tabular}
\label{tab:hp_properties}
\end{table}

\begin{table}[h]
\centering
\caption{Statistical Properties of Kalman-Based Output Gap}
\small
\renewcommand{\arraystretch}{1.5}
\begin{tabular}{|l|c|c|c|c|c|}
\hline
\textbf{Data} & \textbf{Mean} & \textbf{Variance} & \textbf{Skewness} & \textbf{Kurtosis} & \textbf{JB-P Value} \\
\hline
Kalman (Actual) & 0.0227 & 0.0017 & -0.41 & 2.39 & 0.50 \\
Inflation (Actual) & 1.2580 & 0.9721 & -0.22 & 1.83 & 0.35 \\
Simul Inflation & 1.2557 & 0.5879 & -0.72 & 1.95 & 0.10 \\
Simul Output & 0.0273 & 0.0500 & 0.60 & 2.01 & 0.15 \\
\hline
\end{tabular}
\label{tab:kalman_properties}
\end{table}

This confirms that behavioral expectations provide a superior framework for capturing post-pandemic recovery\footnote{For a robust analysis using the whole 20 years of dataset, we find that behavioral expectations consistently outperform rational expectations in matching distributional characteristics. Please refer to the appendix for more details.}, as they align more closely with observed economic dynamics. Standard rational expectations models are insufficient for modeling extreme economic shocks, such as COVID-19, as they fail to account for even the first moment of the output gap and the inflation rate. Additionally, the model reinforces the validity of our approach to model negative aggregate demand and positive aggregate supply shocks within the simple NKDSGE framework. Furthermore, our model quantitatively extends the existing research gap present in the papers of Dasgupta and Rajeev (2020, 2023), where the analysis was static. Moreover, our findings elucidate that vaccination programs generated significant positive supply shocks, counterbalancing the prolonged negative demand shock precipitated by the COVID-19 pandemic. The analysis also reveals that the positive supply shocks' persistence parameter exceeds that of the negative demand shocks, indicating the exceptional efficacy of India's vaccination strategy.

\section{Conclusion}

This study demonstrates that behavioral expectations provide a superior framework for capturing post-pandemic recovery. Behavioral expectations consistently outperform rational expectations in matching distributional characteristics, particularly during periods of extreme economic shock such as the COVID-19 pandemic. Standard rational expectations models are inadequate for modeling such shocks, as they fail to account for even the first moment of key economic variables like the output gap and inflation rate. Our model, which incorporates both negative aggregate demand and positive aggregate supply shocks within the simple NKDSGE framework, further supports the validity of this approach.

Additionally, the study extends the research by Dasgupta and Rajeev (2020, 2023), transitioning from a static model to a dynamic one. The analysis highlights that vaccination programs played a crucial role by generating significant positive supply shocks, which effectively counterbalanced the prolonged negative demand shock caused by the pandemic. Our findings also reveal that the persistence of the positive supply shocks surpasses that of the negative demand shocks, underscoring the exceptional efficacy of India's vaccination strategy in mitigating the pandemic’s economic impact.

% This study contributes to the growing body of literature on macroeconomic modeling for emerging economies by introducing behavioral expectations into a New Keynesian DSGE framework tailored for India. By moving beyond the conventional reliance on rational expectations, we demonstrate that behavioral expectations better capture the moments of India’s output gap data and the inflation rate, particularly in the context of COVID-19-induced economic disruptions. 

% Our quantitative analysis extends previous work by Dasgupta and Rajeev (2023), providing a more rigorous evaluation of the economic impact of India's vaccination program. By modeling the pandemic as a deterministic autoregressive negative demand shock and calibrating the corresponding positive supply shocks, we chose simulated output gap moments to match the moments of the output gap derived from HP and Kalman filters. The results underscore the utility of incorporating behavioral expectations in macroeconomic models to better reflect real-world dynamics.

These findings are particularly relevant for the formation of policy responses to economic crises. The superior performance of the behavioral New Keynesian DSGE model offers policymakers a more accurate and realistic framework. Furthermore, our quantification of vaccination-driven supply shocks provides actionable insights for integrating public health initiatives with economic recovery strategies, contributing to more effective policy design in future crises.

% \subsection{Limitations and Future Directions}
% This analysis is subject to several limitations. First, the absence of actual output gap data for the Indian economy necessitated the construction of a filter based measure using quarterly real GDP data obtained from FRED. Future research should utilize more comprehensive and accurate output gap data for India to enhance the robustness of the results.

% Second, the analysis remained mostly simple behavioral New Keynesian DSGE modelling, where fiscal authority is absent. Future research will extend this model to analyze the role of fiscal policy in analyzing the COVID-19 crisis. 

% Third, this study does not analyze the inflation moments due to the absence of fiscal policy in the model. Fiscal theorists often argue that inflation is fundamentally a fiscal phenomenon, heavily influenced by government spending and debt dynamics. Therefore, we consider it inappropriate to explore inflation-related dynamics within a framework that lacks fiscal policy mechanisms, such as a fiscal rule.

\pagebreak
\section{Appendix}

\subsection{Robustness using a Three Equation Model}
In this sub-section, we check the Jarque-Bera test for the overall 20 years of Output Gap (both HP and Kalman filter-based) and inflation rate data (given in Figure 1 and 2 respectively), establishing that even with larger datasets, the behavioral expectation model captures the distributional characteristics of the data better than the rational expectations model.

To do so, we introduce stochastic autoregressive errors in the three-equation New Keynesian (NK) model (described in our model section), with a persistence parameter of 0.95 and Gaussian error with a mean of 0 and variance of 0.5. We then obtain a simulated snapshot of 81 datapoints from the overall simulated data of 2000 sample size. The simulation is taken from observations 1000 to 1080. 

The results of the Jarque-Bera tests for normality of the data are summarized below:

\begin{table}[h!]
\centering
\caption{Comparison of Distributions for Expectation Formation Models and Real Data}
\small
\renewcommand{\arraystretch}{1.1}
\begin{tabular}{|l|c|c|}
\hline
\textbf{Data Type}                 & \textbf{Normality (5\%)} & \textbf{p-value} \\ \hline
Simulated Behavioral Output Gap    & Non-Normal              & 0.03            \\
Actual Output Gap (Kalman)         & Non-Normal              & \textless{}0.001 \\
Simulated Rational Output Gap      & Normal                  & 0.08            \\
Actual Output Gap (HP)             & Non-Normal              & \textless{}0.001 \\
Simulated Behavioral Inflation Rate & Non-Normal              & 0.02            \\
Simulated Rational Inflation Rate  & Normal                  & 0.20            \\
Actual Inflation Rate              & Non-Normal              & 0.00597         \\ \hline
\end{tabular}
\label{table:expectation_comparison}
\end{table}

% \begin{table}[h!]
% \centering
% \caption{Comparison of Distributions for Expectation Formation Models and Real Data}
% \begin{tabular}{@{}lcccc@{}}
% \toprule
% \textbf{Data Type}                 & \textbf{Normality (5\%)} & \textbf{p-value} \\ \midrule
% Simulated Behavioral Output Gap    & Non-Normal              & 0.03            \\
% Actual Output Gap (Kalman)         & Non-Normal              & \textless{}0.001 \\
% Simulated Rational Output Gap      & Normal                  & 0.08            \\
% Actual Output Gap (HP)             & Non-Normal              & \textless{}0.001 \\
% Simulated Behavioral Inflation Rate & Non-Normal              & 0.02            \\
% Simulated Rational Inflation Rate  & Normal                  & 0.20            \\
% Actual Inflation Rate              & Non-Normal              & 0.00597         \\ \bottomrule
% \end{tabular}
% \label{table:expectation_comparison}
% \end{table}
Even with the full dataset, we observe that the behavioral expectation model better matches the distributional characteristics of both the Indian output gap and the quarterly inflation rate. This result replicates the findings of De Grauwe (2012)\footnote{De Grauwe (2012) did the analysis using FRED data of the USA and found similar results for the US economy. Here, we replicate his fundamental contribution for India.} and further justifies our results with enhanced robustness.

\subsection{Appendix: estimation of the Output gap}

To compute the output gap\footnote{See, Goyel and Arora (2016) for a similar analysis using the HP and Kalman filter to estimate the potential GDP of India.}, this paper employs two techniques: (i) the Hodrick-Prescott (HP) filter, and, (ii) Dynamic Linear Models based on the Kalman filter, where the output gap is as follows,
\[
\text{Output Gap} = \log(\text{Observed GDP}) - \log(\text{Potential GDP}),
\]

\subsubsection{Hodrick-Prescott Filter}
The Hodrick-Prescott filter minimizes an objective function to decompose logarithm of real GDP (\(z_t\)) into its trend component (\(t_t\)) and cyclical component. The filter actively balances two competing goals: fitting the observed GDP as closely as possible and ensuring the smoothness of the trend. The objective function is as follows:
\[
\min \sum_{t=1}^n (z_t - t_t)^2 + \lambda \sum_{t=2}^{n-1} \left[(z_{t+1} - t_t) - (t_t - t_{t-1})\right]^2,
\]
where the first term measures the goodness of fit by minimizing the deviations of the log real GDP (\(z_t\)) from the trend (\(t_t\)), and the second term penalizes changes in the growth rate of the trend to ensure smoothness. This paper sets \(\lambda = 1600\), as we have quarterly data. We used the data of the detrended output gap in the output gap (HP).

\subsubsection{Kalman Filter}
Dynamic linear models often consist of two main equations: the \textit{measurement equation} and the \textit{state evolution equation}.
% \subsubsection{Measurement Equation}
The measurement equation relates the logarithm of observable GDP (\(z_t\)) to the unobservable state variables (\(\mathbf{s}_t\)), which include the logarithm of potential GDP (\(s_{1t}\)) and the trend growth rate (\(s_{2t}\)):
\[
z_t = \mathbf{F} \mathbf{s}_t + \nu_t,
\]
where,
\[
\mathbf{F} = [1, 1],
\]
\[
\mathbf{s}_t = 
\begin{bmatrix}
s_{1t} \\
s_{2t}
\end{bmatrix},
\]
and \(\nu_t \sim \mathcal{N}(0, V)\) is the measurement noise with variance \(V = 0.06^2\).

% \subsubsection{State Evolution Equation}
The state evolution equation describes how the state variables evolve over time:
\[
\mathbf{s}_{t+1} = \mathbf{G} \mathbf{s}_t + \omega_t,
\]
where,
\[
\mathbf{G} = 
\begin{bmatrix}
1 & 0 \\
1 & 1
\end{bmatrix},
\]
and \(\omega_t \sim \mathcal{N}(0, \mathbf{W})\) represents the process noise with covariance matrix:
\[
\mathbf{W} = 
\begin{bmatrix}
0.06^2 & 0 \\
0 & 0.06^2
\end{bmatrix}.
\]
% \subsubsection{Initial State and Covariance}
Note, the model starts with an initial guess for the state variables,
\[
\mathbf{s}_0 = 
\begin{bmatrix}
\log(\text{Observed GDP at } t=1) \\
0
\end{bmatrix},
\]
and assumes an initial covariance matrix (See; Goyel and Arora, 2016)
\[
\mathbf{C}_0 = 
\begin{bmatrix}
0.06^2\ & 0 \\
0 & 0.06^2\
\end{bmatrix}.
\]

% \subsubsection{Kalman Filter Steps}
The model then uses the Kalman filter to estimate the parameters and states recursively. The filter alternates between the following steps:

\paragraph{Prediction Step}
Predict the state and its covariance for time \(t+1\):
\[
\hat{\mathbf{s}}_{t+1|t} = \mathbf{G} \hat{\mathbf{s}}_{t|t},
\]
\[
\mathbf{P}_{t+1|t} = \mathbf{G} \mathbf{P}_{t|t} \mathbf{G}^\top + \mathbf{W},
\]
where \(\hat{\mathbf{s}}_{t+1|t}\) represents the predicted state and \(\mathbf{P}_{t+1|t}\) represents its covariance.

\paragraph{Update Step}
Kalman filter incorporates new observations to refine the state estimate:
\[
\mathbf{K}_t = \mathbf{P}_{t+1|t} \mathbf{F}^\top (\mathbf{F} \mathbf{P}_{t+1|t} \mathbf{F}^\top + V)^{-1},
\]
\[
\hat{\mathbf{s}}_{t+1|t+1} = \hat{\mathbf{s}}_{t+1|t} + \mathbf{K}_t \left( z_{t+1} - \mathbf{F} \hat{\mathbf{s}}_{t+1|t} \right),
\]
\[
\mathbf{P}_{t+1|t+1} = (\mathbf{I} - \mathbf{K}_t \mathbf{F}) \mathbf{P}_{t+1|t}.
\]
Here, \(\mathbf{K}_t\) represents the Kalman gain, \(\hat{\mathbf{s}}_{t+1|t+1}\) represents the updated state estimate, and \(\mathbf{P}_{t+1|t+1}\) represents its covariance.

% \subsubsection{Output Gap Estimation}
Finally we obtain the output gap (\(y_t\)) using\footnote{The R code will be made available on request. Please note that this code truncates the first observation of the estimated logarithm of potential GDP as the first two estimates become identical.}:
\[
\text{Output Gap}_t = (z_t) - (s_{1t}),
\]
where \(z_t\) represents the logarithm of observed GDP and \(s_{1t}\) represents the estimated logarithm of potential GDP at time \(t\).
\subsection{Appendix: Structural break}
We evaluate the structural break at around Q1 2020 using the Likelihood ratio test. The test evaluates two nested models:

\begin{itemize}
    \item \textbf{Model 1 (Null Hypothesis)}: Assumes no structural break and fits a constant mean model:
    \[
    Y_t = \mu + \epsilon_t,
    \]
    where \(Y_t\) is the observed output gap, \(\mu\) is the constant mean, and \(\epsilon_t\) represents the error term.
    \item \textbf{Model 2 (Alternative Hypothesis)}: Incorporates a structural break at \textbf{Q1 2020}, dividing the dataset into pre- and post-break periods:
    \[
    Y_t = \mu_1 I(t \leq \text{Q1 2020}) + \mu_2 I(t > \text{Q1 2020}) + \epsilon_t,
    \]
    where \(\mu_1\) and \(\mu_2\) represent the means for the pre and post-break periods, respectively, and \(I\) is an indicator function.
\end{itemize}

The \textit{Analysis of Variance (ANOVA)} results for the HP and Kalman filters are presented below.

\begin{table}[h!]
\centering
\caption{ANOVA Table for Likelihood Ratio Test (HP Filter).}
\small
\renewcommand{\arraystretch}{1.1}
\begin{tabular}{|l|c|c|}
\hline
\textbf{Metric} & \textbf{Model 1 (No Break)} & \textbf{Model 2 (With Break)} \\ \hline
\textbf{Residual Degrees of Freedom (Df)} & 80 & 79 \\
\textbf{Residual Sum of Squares (RSS)} & 0.24 & 0.21 \\
\textbf{Degrees of Freedom (Df)} & --- & 1 \\
\textbf{Sum of Squares (SS)} & --- & 0.03 \\
\textbf{F-statistic} & --- & 11.52 \\
\textbf{p-value (\(Pr(>F)\))} & --- & 0.001 \\ \hline
\end{tabular}
\label{tab:lr_test_hp}
\end{table}

% \begin{table}[h!]
% \centering
% \caption{ANOVA Table for Likelihood Ratio Test (HP Filter).}
% \label{tab:lr_test_hp}
% \begin{tabular}{lcc}
% \hline
% \textbf{Metric} & \textbf{Model 1 (No Break)} & \textbf{Model 2 (With Break)} \\
% \hline
% \textbf{Residual Degrees of Freedom (Df)} & 80 & 79 \\
% \textbf{Residual Sum of Squares (RSS)} & 0.24 & 0.21 \\
% \textbf{Degrees of Freedom (Df)} & --- & 1 \\
% \textbf{Sum of Squares (SS)} & --- & 0.03 \\
% \textbf{F-statistic} & --- & 11.52 \\
% \textbf{p-value (\(Pr(>F)\))} & --- & 0.001 \\
% \hline
% \end{tabular}
% \end{table}

Tables III and IV present the Likelihood Ratio test results for the HP and Kalman filter-derived output gap series, respectively. For the HP-filtered series (Table III) (or Model 2), which incorporates a Q1 2020 structural break, demonstrates superior fit with a reduced Residual Sum of Squares (RSS) of 0.21 compared to Model 1's RSS of 0.24. The associated F-statistic of 11.52 confirms the statistical significance of this structural break.
\begin{table}[h!]
\centering
\caption{ANOVA Table for Likelihood Ratio Test (Kalman Filter).}
\small
\renewcommand{\arraystretch}{1.1}
\begin{tabular}{|l|c|c|}
\hline
\textbf{Metric} & \textbf{Model 1 (No Break)} & \textbf{Model 2 (With Break)} \\ \hline
\textbf{Residual Degrees of Freedom (Df)} & 80 & 79 \\
\textbf{Residual Sum of Squares (RSS)} & 0.13 & 0.09 \\
\textbf{Degrees of Freedom (Df)} & --- & 1 \\
\textbf{Sum of Squares (SS)} & --- & 0.03 \\
\textbf{F-statistic} & --- & 28.41 \\
\textbf{p-value (\(Pr(>F)\))} & --- & \(9.06 \times 10^{-7}\) \\ \hline
\end{tabular}
\label{tab:lr_test_kalman}
\end{table}

% \begin{table}[h!]
% \centering
% \caption{ANOVA Table for Likelihood Ratio Test (Kalman Filter).}
% \label{tab:lr_test_kalman}
% \begin{tabular}{lcc}
% \hline
% \textbf{Metric} & \textbf{Model 1 (No Break)} & \textbf{Model 2 (With Break)} \\
% \hline
% \textbf{Residual Degrees of Freedom (Df)} & 80 & 79 \\
% \textbf{Residual Sum of Squares (RSS)} & 0.13 & 0.09 \\
% \textbf{Degrees of Freedom (Df)} & --- & 1 \\
% \textbf{Sum of Squares (SS)} & --- & 0.03 \\
% \textbf{F-statistic} & --- & 28.41 \\
% \textbf{p-value (\(Pr(>F)\))} & --- & \(9.06 \times 10^{-7}\) \\
% \hline
% \end{tabular}
% \end{table}

The Kalman-filtered series (Table IV) exhibits even stronger evidence of a structural break. The RSS decreases substantially from 0.13 in Model 1 to 0.09 in Model 2, yielding a higher F-statistic of 28.41. Both filtering methodologies thus provide robust empirical support for a structural break in India's output gap at Q1 2020.

\end{document}